\documentclass[preprint,preprintnumbers,superscriptaddress,amsmath,amssymb,floatfix,prl]{revtex4-1}
\usepackage{graphicx,color}           % Include figure files

\def\nm{{\ {\rm nm}}}                       % nm
\def\gauss{{\ {\rm G}}}                     % gauss
\def\Hz{{\ {\rm Hz}}}                       % Hz
\def\kHz{{\ {\rm kHz}}}                     % kHz
\def\MHz{{\ {\rm MHz}}}                     % MHz
\def\Er{E_R}                            % Er
\def\kr{k_R}                            % kr
\def\Rb87{^{87}\text{Rb}}                     % Rb 87
\def\Na23{^{23}\text{Na}}                     % Na 23
\def\K41{^{41}\text{K}}                     % Na 23
\def\Li6{^{6}\text{Li}}                       % Li 6
\def\ex{{\mathbf e}_x}                            % e_x
\def\ey{{\mathbf e}_y}                            % e_y
\def\ez{{\mathbf e}_z}                            % e_z
\def\arf{\Omega_{\text{rf}}}
\def\trf{\text{rf}}
\def\xibb{\boldsymbol{\xi}_{\beta,\beta'}}
\def\xirf{\boldsymbol{\xi}_{\trf}}

\DeclareMathAlphabet\mathbfcal{OMS}{cmsy}{b}{n}
\def\bra#1{\mathinner{\langle{#1}|}}
\def\ket#1{\mathinner{|{#1}\rangle}}

{\catcode`\|=\active
  \gdef\Braket#1{\left<\mathcode`\|"8000\let|\BraVert {#1}\right>}}
\def\BraVert{\egroup\,\mid@vertical\,\bgroup}

\begin{document}

\title{Rashba realization: Raman with RF}

\author{D.~L.~Campbell and I.~B.~Spielman}
\address{Joint Quantum Institute, University of Maryland and National Institute of Standards and Technology, College Park, Maryland, 20742, USA}
%\ead{dlcamp@umd.edu}
%\ead{ian.spielman@nist.gov}

\date{\today}

\begin{abstract}

We theoretically explore a Rashba spin-orbit coupling scheme which operates entirely in the absolute ground state manifold of an alkali atom, thereby minimizing all inelastic processes.  An energy gap between ground eigenstates of the proposed coupling can be continuously opened or closed by modifying laser polarizations.  Our technique uses far-detuned ``Raman" laser coupling to create the Rashba potential, which has the benefit of low spontaneous emission rates.  At these detunings, the Raman matrix elements that link $m_F$ magnetic sublevel quantum numbers separated by two are also suppressed.  These matrix elements are necessary to produce the Rashba Hamiltonian within a single total angular momentum $f$ manifold.  However, the far-detuned Raman couplings can link the three XYZ states familiar to quantum chemistry, which possess the necessary connectivity to realize the Rashba potential.  We show that these XYZ states are essentially the hyperfine spin eigenstates of $^{87}\text{Rb}$ dressed by a strong radio-frequency magnetic field.

\end{abstract}


\maketitle

\section{introduction}

Geometric gauge potentials are encountered in many areas of physics~\cite{Berry1984,Jackiw1988,Moody1986,Bohm1992,Zee1988,Shapere1989,Mead1992,Bohm2003,Xiao2010}.  In atomic gases, the geometric vector and scalar potentials were first considered in the late 90's to fully describe atoms ``dressed'' by laser beams~\cite{Dum1996,Visser1998,Dutta1999}.  Atoms that move in a spatially varying, internal state dependent optical field experience geometric vector and scalar potentials.  Our understanding of these potentials has been refined, and now allow for the engineered addition of spatially homogeneous geometric gauge potentials~\cite{Lin2011, Wang2012, Cheuk2012}.  In many cases, the resulting atomic Hamiltonian is equivalent to iconic models of spin-orbit coupling (SOC): Rashba, Dresselhaus and combinations thereof.  

% in addition to the energies associated with the local eigenenergies

Often, systems with spin-orbit coupling will have multiply degenerate single particle eigenstates with topological character: this suggests that strongly correlated phases will exist in the presence of interactions for both bosonic and fermionic systems.  Interesting phenomena such as topological insulating states and the spin-Hall effect include SOC as a necessary component~\cite{Kane2005, Hasan2010}.  Rashba SOC (present for 2D free electrons in the presence of a uniform perpendicular electric field, such as in asymmetric semiconductor heterostructures)~\cite{Bychkov1984, Shen2004}, is an iconic 2D SOC potential and has maximal ground state symmetry.  Indeed, interesting many-body phases~\cite{Jian2011,Zhang2012,Larson2009} predicted for atomic systems with Rashba SOC include unconventional and fragmented Bose-Einstein condensation~\cite{Stanescu2008}, composite fermion phases of bosons~\cite{Sedrakyan2012} and anisotropic or topological superfluids in fermionic systems~\cite{Hu2011}.

It is in the context of such potentially fragile many-body states that we propose a scheme that is implemented entirely within the ground hyperfine manifold of an alkali with spin greater than or equal to spin-1.  Recently,  the Rashba potential was realized with $40 \text{K}$ fermions using lasers coupling the $f=7/2$ and $f=9/2$ manifolds~\cite{Huang2015}.  In alkali bosonic systems with density $n$ the 2-body collisional relaxation lifetime from the $f+1$ to the $f$ ground state hyperfine manifold is $>n\times 10^{-14}\text{ cm}^3/{\rm s}$~\cite{Tojo2009}: a timescale that is potentially too small to observe meaningful many-body physics.  Such relaxation may be a lesser, but still pertinant, concern in fermionic systems.  We propose an alternative coupling scheme implemented entirely within the ground hyperfine manifold of alkali atoms.  

\subsection{Rashba SOC for electrons}

\begin{figure}
\begin{centering}
\includegraphics[width=6.0in]{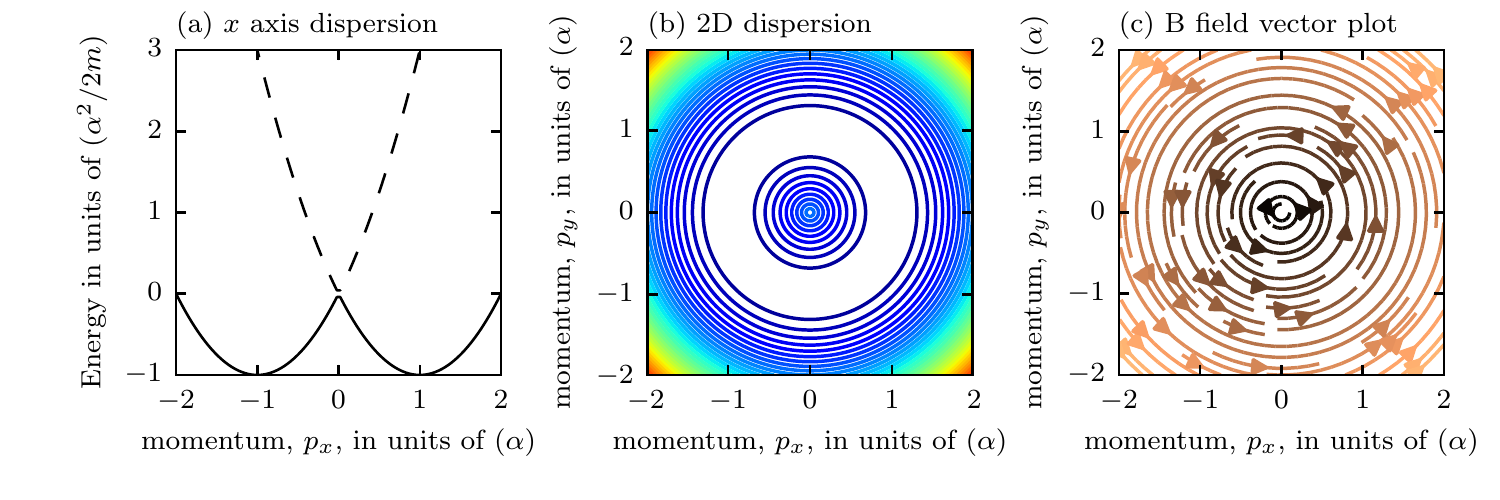}
\caption[Vector plot, dispersion and contour]{\label{fig:IdealRashba} Rashba dispersion in electron systems. (a) Cross-sectional cut of the 2D Rashba dispersion  (b) Contour plot of dispersion demonstrating cylindrical symmetry.  (c)  Vector plot of ${\bf B}_{{\rm SOC}} = \hbar{\bf k}/m\times {\bf E}/c^2$: the ground state electron spin is antialigned with ${\bf B}$.   An electron that loops about the momentum origin adiabatically traces out the equator on the Bloch sphere.  After one full loop a Berry's phase of $\pi$ is acquired.}
\end{centering}
\end{figure} 

The simplest model of Rashba SOC describes a 2D free electron system in terms of electron momentum $\hbar {\bf k}$ and gyromagnetic ratio $g$ in the presence of an out-of-plane electric field ${\bf E} = E \ez$.  We consider the electrons relativistically: in the electron's moving frame an in-plane magnetic field ${\bf B}_{{\rm SOC}} = \hbar{\bf k}/m\times {\bf E}/c^2$ appears in proportion to momentum, as shown in Fig.~(\ref{fig:IdealRashba}).  The additional contribution to the spin-1/2 electron's Zeeman Hamiltonian from ${\bf B}_{{\rm SOC}}$ is 
\begin{align}
\hat{H}_{{\rm SOC}}=\frac{2\alpha}{m}({\bf k}\times \ez)\cdot  \hbar\hat{\boldsymbol{\sigma}}/2,
\end{align}
where $\alpha = g\mu_B|{\bf E}|/2c^2$, $\hbar\hat{\boldsymbol{\sigma}}/2$ is the electron spin operator, and $\hat{\boldsymbol{\sigma}}=(\hat\sigma_x, \hat\sigma_y, \hat\sigma_z)$ is the vector of Pauli matrices.  As shown in Fig.~(\ref{fig:IdealRashba})ab, a degenerate ring of momenta described by $k_x^2+k_y^2 = \alpha^2$ comprises the ground state of this Hamiltonian.  At the origin (${\bf k}=0$) the eigenenergies of the Rashba Hamiltonian intersect: this point is often called a Dirac point.

Ignoring overall energy shifts, the Hamiltonian including $\hat{H}_{{\rm SOC}}$ and the kinetic energy can be expressed as $\hat{H}=(\hbar{\bf k}-\hat{\mathcal{A}})^2/2m$, in terms of an effective vector potential $\hat{\mathcal{A}}=\alpha(\hat \sigma_y\ex-\hat \sigma_x\ey)$.  The Cartesian components of the vector potential manifestly fail to commute: the vector potential is non-abelian.

An atom that adiabatically traverses a loop about the momentum origin in Fig.~(\ref{fig:IdealRashba})bc acquires a Berry's phase of $\pi$.  An interferometer in which one arm traverses the momentum origin would display destructive interference.  It is anticipated that the presence of this phase winding will result in unusual many-body ground states for both fermionic and bosonic systems~\cite{Stanescu2008, Sedrakyan2012, Hu2011}. 

%We parametrize the two $\pm$ eigenstates of $\hat H$ with the electron momentum $\hbar{\bf k} = \hbar(k_x,k_y)$: $\ket{{\bf k}, \pm}$.  $\mathcal{Q}$ is a set of points that defines a loop in $\hbar {\bf k}$.  We may determine the acquired Berry's phase for a single traversal of an electron through all the points in $\mathcal{Q}$
%%
%\begin{equation}
%\gamma_{\pm} = \oint_\mathcal{Q} d\mathcal{Q} \bra{\mathcal{Q},\pm}\boldsymbol{\nabla}_{\mathcal{Q}} \ket{\mathcal{Q}, \pm}
%\end{equation}
%%
%which is $\pm\pi$ for $\mathcal{Q}$ that encircle ${\bf k}=0$ and zero otherwise.  

\section{Rashba SOC in cold atoms}

\subsection{Overview}

\begin{figure}
\begin{centering}
\includegraphics[width=6.2in]{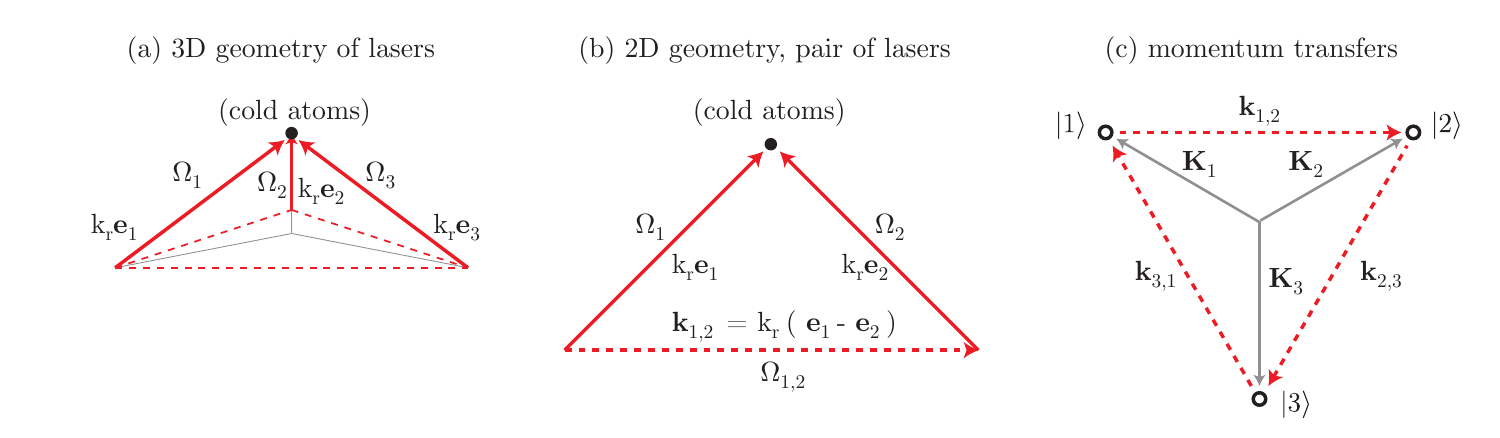}
\caption[Rashba coupling sketch]{Schematic view of laser geometry necessary for SOC of three states.  (a) We show an example three beam laser geometry where the beam wavevectors $\kr {\bf e}_{j}$, for $j\in\{1,2,3\}$ indexing beams, are all mutually orthogonal.  (b) A 2D cutout of two beams shows that the magnitude of the difference between the wavevectors of beams 1 and 2 is $\sqrt{2}\kr$.  In panel (c), we show a self consistent picture for the momentum recoil transferred $\hbar k_{j,j'}$ by a Raman excitation that changes the atomic eigensate from $\ket{j'}$ to $\ket{j}$.\label{fig:sketch}}
\end{centering}
\end{figure} 

In ultracold neutral atoms SOC is produced by coupling the atomic ground states with laser fields, e.g. two-photon Raman coupling, where the laser fields impart a discrete momentum kick whenever they induce a spin flip.  We consider a subspace of three long-lived states $\ket{j}\in\{1,2,3\}$ within a potentially much larger pool of available states.  We illuminate these states with three coherent lasers that are indexed by $\beta\text{, }\beta'\in\{1,2,3\}$.  Each of these lasers has distinct wavevector ${\bf k}_{\beta}$, magnitude $\kr$, and frequency 
\begin{align}
f_{\beta} = (\omega_L-\omega_{\beta})/2\pi\label{eq:freq}.
\end{align}

The three possible two-photon Raman frequencies differences are given by $\omega_{\beta, \beta'} = -(\omega_{\beta} - \omega_{\beta'})$.  Likewise, there are three distinct difference wavevectors ${\bf k}_{\beta, \beta'} = {\bf k}_{\beta}-{\bf k}_{\beta'}$ and three distinct difference phases between beams $\gamma_{\beta,\beta'} = \gamma_{\beta}-\gamma_{\beta'}$.  Figure~(\ref{fig:sketch}) illustrates the relationship between laser momentum recoil $\hbar {\bf k}_{\beta}$, with magnitude $\hbar \kr$, and Raman recoil $\hbar {\bf k}_{\beta, \beta'}$.  The Hamiltonian describing Raman coupling in this general form is
\begin{align}
 \hat H({\bf k}) = &\sum_{j, j'}\bigg\{\left[\frac{\hbar^2{\bf k}^2}{2m}+E_{j}\right] \delta_{j,j'} \label{eq:Hx}\\
+&\sum_{\beta\ne\beta'}\hbar \Omega_{j,j',\beta,\beta'} \exp{(i [{\bf k}_{\beta, \beta'}\cdot {\bf x}-\omega_{\beta,\beta'}t-\gamma_{\beta,\beta'}])}(1-\delta_{j,j'})\bigg\}\ket{j}\bra{j'}\nonumber
\end{align}
where $E_{j}$ is the eigenenergy of state $\ket{j}$ in the absence laser-coupling.  

We shall make the simplifying assumption that each pair of Raman lasers uniquely couples a pair of states, greatly simplifying the form of the coupling amplitude in Eq.~(\ref{eq:Hx}): $\Omega_{j,j'}$.  This configuration can be relaized by requiring that the Raman lasers resonantly couple pairs of states
\begin{align}
\hbar\omega_{j,j'}=E_{j}-E_{j'}\label{eq:resonance}\\
\hbar \omega_j = E_j
\end{align}
where we have linked each Raman beam to a state with this resonance condition (recall that the Raman laser frequencies are given by Eq.~\ref{eq:freq}).  We also apply the rotating wave approximation (RWA) to eliminate terms that are $\propto\exp{(i\omega_{j,j'}t)}\ket{j}\bra{j'}$.

With these constraints on Eq.~(\ref{eq:Hx}) it is always possible to apply a unitary transformation that eliminates the complex exponentials from the Hamiltonian
\begin{align}
\hat{U}({\bf x}, t) = \sum_{j}\exp{(i[{\bf k}_{j}\cdot {\bf x}-\omega_{j}t-\gamma_{j}])}\ket{j}\bra{j}\label{eq:unitarytrans},
\end{align}
and also applies a state-dependent momentum displacement to the momentum operator in Eq.~(\ref{eq:Hx}).  In the rotating frame, $\hat{U}({\bf x}, t) \hat H({\bf k})\hat{U}^{\dagger}({\bf x}, t)$ is
\begin{align}
\hat{H}({\bf q})= \sum_{j,j'}\bigg\{&\frac{-\hbar^2 ({\bf q}-{\bf k}_{j})^2}{2m}\delta_{j,j'}\label{eq:Hq}+\hbar \Omega_{j,j'}(1-\delta_{j,j'})\bigg\}\ket{j}\bra{j'}
\end{align}
where the matrix element $\Omega_{j,j'}$ is potentially complex.

%but not without modifying the diagonal elements of Eq.~(\ref{eq:Hx}).  The position dependent term in $\hat{U}({\bf x}, t)$ is a state-dependent momentum displacement operator $\hat{U}({\bf x}, t){\bf k}^2\hat{U}^{\dagger}({\bf x}, t) = \sum_j ({\bf q}-{\bf k}_j)^2\ket{j}\bra{j}$.

\subsection{Rashba subspace}

We apply a discrete Fourier transform
\begin{align}
\ket{n} = \frac{1}{\sqrt{3}}\sum_{j=1}^{3} \exp{(-i 2 \pi j n / 3 )} \ket{j}.
\end{align}
to Eq.~(\ref{eq:Hq}).  This is a useful diagonalization tool when all the off-diagonal matrix elements are nearly equal in amplitude and larger than the energy scale of any of the three two photon recoils $\hbar^2 {\bf k}_{j,j'}^2/2m$.  We specify our discussion to equal amplitudes $\Omega=|\Omega_{j,j'}|$ for each matrix element.  We also define a phase $\phi_{j,j'} = i\ln{(\Omega_{j,j'}/|\Omega_{j,j'}|)}$.  In the transformed basis the eigenenergies of the atom-light interaction are 
\begin{align}
E_n &= 2\hbar\Omega\cos{(2\pi n/3+\bar\phi)}\label{eq:Energy}\\
\bar \phi &= (\phi_{3,2}+\phi_{2,1}+\phi_{1,3})/3 =  -(\phi_{2,3}+\phi_{1,2}+\phi_{3,1})/3.
\end{align}
The phase sum $\bar \phi$ adds the phase contributions from nearest neighbor matrix elements that sequentially chains all three states together.  $\bar\phi$ is an example of a phase that is not simply the result of our choice of basis: it cannot be eliminated by the transformation in Eq.~(\ref{eq:unitarytrans}).   If $\bar\phi = 0$ the states $\ket{n=1}$ and $\ket{n=2}$ are degenerate in energy.

We define an effective vector ${\bf k}_{j,j'} = {\bf K}_j-{\bf K}_{j'}$ where ${\bf K}_j = {\bf k}_j-\sum_{j=1}^3 {\bf k}_j$.  
When ${\bf K}_{j}={\rm k}_{{\rm eff}}[\cos{(2\pi j/3)}\ex+\sin{(2\pi j/3)}\ey]$ are the vertices of an equilateral triangle the Hamiltonian in the discrete Fourier basis is
\begin{align}
\hat{H}=&\sum_{n=1}^{3}\left[\frac{\hbar^2{\bf q}^2}{2m}-\hbar\Omega\cos{(2\pi n/3+\bar{\phi})}\right]\ket{n}\bra{n}\label{eq:Rashba}\\
+&\frac{\hbar^2 {\rm k}_{{\rm eff}}^2}{m}\left[(iq_x+q_y)\right(\ket{1}\bra{3}+\ket{3}\bra{2}+\ket{2}\bra{1}\left)+\text{h.c.}\right]\nonumber.
\end{align}

We neglect the most energetic state when $\bar \phi = 0$ and recover the two-state Rashba Hamiltonian
\begin{align}
\hat{H}_{\text{sub}} = \frac{\hbar^2\boldsymbol{q}^2}{2m}\hat{1}+\frac{\hbar^2 {\rm k}_{{\rm eff}}}{m}(\hat\sigma_x q_y-\hat \sigma_y q_x)+\hat\sigma_z \bar\phi\label{eq:twostateRashba}
\end{align}
where $\hat{1}$ is the identity for a two state system.  The last term in Eq.~(\ref{eq:twostateRashba}) describes a gap opening at ${\bf q}=0$ between the ground eigenstates for small values for $\bar\phi$.  

\subsection{Physical implementation and limitations}

\begin{figure}
\begin{centering}
\includegraphics[width=5.0in]{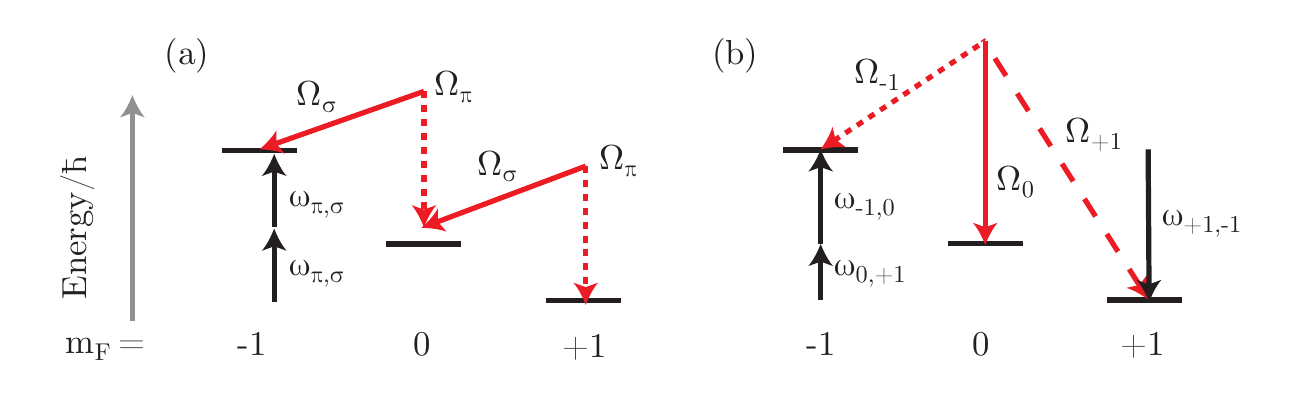}
\caption[Quasimomentum coupling geometry and schematic levels]{Schematic view of spin-orbit coupling in the spin-1 ground state of $\Rb87$.  (a) In most current experiments, the same Raman lasers simultaneously couple pairs of adjacent spin states.  Momentum transfers in the quasimomentum basis are arranged in a line.  (b) For Rashba SOC, such as the realization of Ref.~\cite{Huang2015}, all pairs of spin states are independently coupled and momentum transfers in the quasimomentum basis are the vertices of an equilateral triangle.  \label{fig:levels}}
\end{centering}
\end{figure} 

As made evident by its presence in Eq.~(\ref{eq:Hx}) and absence in Eq.~(\ref{eq:Hq}) the phase of each Raman beam does not contribute to the steady state Hamiltonian.  This symmetry is absent when there are more than three Raman frequency differences for a three state subsystem or in ring coupling geometries with $N>3$~\cite{Campbell2011}.  This consideration is very compelling from an experimental perspective since small variations in the pathlength of each laser could otherwise produce dramatic changes in the potential.

Direct Raman coupling of the hyperfine ground states of alkali atoms using far off-resonant coupling of the electronic excited states cannot couple states with an angular momentum difference greater than $1$ unit of angular momentum.  Coupling as shown in Fig.~(\ref{fig:levels})a is possible, while coupling as shown in Fig.~(\ref{fig:levels})b is not.  Detuning near the excited electronic hyperfine states as proposed in Ref.~\cite{Ruseckas2005} lifts the angular momentum restriction sufficiently to realize a coupling scheme similar to Fig.~(\ref{fig:levels})b but the spontaneous emission rate increases and atomic ensemble lifetimes become much shorter than typical equilibration times.  Three states with sufficient connectivity to produce the Rashba Hamiltonian can be obtained by including states from multiple ground electronic $f$ manifolds~\cite{Juzeliunas2010}.  The matrix elements that link these states are proportional to the detuning $1/\Delta = 1/\Delta_{3/2}-1/\Delta_{1/2}$ from the $P_{3/2}$ and $P_{1/2}$ lowest electronic fine structure.  A phase of $\pi$ is contributed to $\bar \phi$ when $1/\Delta$ is negative and $0$ otherwise.  Recently, an experiment realized the Rashba dispersion using positive $1/\Delta$~\cite{Huang2015}.  Although feasable, collisions that change $f$ are expected to lead to atom-loss and heating, potentially decohering fragile many-body phases.  Here, we detail a scheme that produces the Rashba Hamiltonian entirely in the $f=1$ ground state manifold of $\Rb87$.

\section{Physical Implementation}

\subsection{Form of the Raman coupling} 

We introduce the local electric field ${\bf E}(t)=\sum_{\beta} {\bf E}_{\beta} \cos{({\bf k}_{\beta}\cdot{\bf x}-\omega_{\beta} t-\gamma_{\beta})}$ of linearly polarized lasers impinging upon an atomic system.  This gives a coupling
\begin{align}
\hat{H}_{{\rm eff}} &= \frac{g_F \mu_B}{\hbar}{\bf B}_{{\rm eff}}\cdot \hat {\bf F}\label{eq:Heff0th}\\
{\bf B}_{{\rm eff}} &= \frac{i u}{g_S\mu_B}({\bf E}^{*}({\bf x}, t)\times{\bf E}({\bf x}, t))\label{eq:Beff},
\end{align}
in the ground hyperfine manifold of an alkali atom, where $g_S$ is the gyromagnetic ratio of the electron spin, $g_F$ is the Land\'e g-factor for the hyperfine states, $\mu_B$ is the Bohr magneton, and the two-photon vector light shift matrix element is
\begin{align}
u = \frac{|\langle||{\bf d}||\rangle|^2}{4}\left(\frac{1}{3\Delta_{3/2}}-\frac{1}{3\Delta_{1/2}}\right).
\end{align}
The far off-resonant Wigner-Eckart reduced matrix element is given by $\langle||{\bf d}||\rangle = \langle l=0||{\bf d}|| l=1 \rangle$ where $l=0,1$ is the orbital angular momentum quantum number for the ground and excited electronic states, respectively.

We compute the pairwise product of components of the local electric field in the effective Hamiltonian Eq.~(\ref{eq:Heff0th}) and retain terms that have
\begin{align}
\Phi_{\beta,\beta'} = {\bf k}_{\beta,\beta'}\cdot {\bf x}-\omega_{\beta,\beta'} t-\gamma_{\beta,\beta'}
\end{align}
in the argument of the complex exponentials.

When laser polarizations are linear we may rearrange terms and obtain the effective coupling between the ground electronic hyperfine spin projections
\begin{align}
\hat{H}_{{\rm eff}} &= \sum_{\beta\ne\beta'}\frac{-g_F u |{\bf E}_{\beta} \times {\bf E}_{\beta'}|}{2\hbar g_S}\sin{[\Phi_{\beta,\beta'}]}\cdot \hat{{\bf F}},\label{eq:Heff1st}\\
\end{align}
where
\begin{align}
\hat {\bf F} = (\hat F_x, \hat F_y, \hat F_z)
\end{align}
is the vector of spin-1 operators.

We tune some Raman frequency differences to near resonance $|\omega_{\beta,\beta'}-\delta_Z| \ll \delta_Z$ with the linear Zeeman splitting $\hbar\delta_Z = g_F\mu_B B_{{\rm dc}}$ produced by a dc magnetic field $B_{{\rm dc}}\ez$.  We apply a RWA to these and keep couplings proportional to $\hat{F}_{x,y}$
\begin{align}
\hat{H}^{\perp}_{\beta,\beta'} = \Omega^{\perp}_{\beta,\beta'} \hat F_+ e^{i \Phi_{\beta,\beta'}}\theta(\omega_{\beta,\beta'})+\text{h.c.}
\end{align}
where $\theta$ is the Heaviside function and $\hat F_+ = \hat F_x+i\hat F_y$.  The matrix elements $\Omega^{\perp}_{\beta,\beta'}$ in the RWA are
\begin{align}
\Omega^{\perp}_{\beta,\beta'} &= \frac{i g_F u |{\bf E}_{\beta} \times {\bf E}_{\beta'}|}{4\hbar g_S}\xibb\cdot (\ex+i\ey)\label{eq:omgperp}
\end{align}
where
\begin{align}
\xibb &= \frac{{\bf E}_{\beta} \times {\bf E}_{\beta'}}{|{\bf E}_{\beta} \times {\bf E}_{\beta'}|}
\end{align}
are complex unitary numbers that take on different values when the vector orientations of ${\bf E}_{\beta} \times {\bf E}_{\beta'}$ differ but the same pair of hyperfine spin projections are coupled.  The Hamiltonian in Eq.~(\ref{eq:Heff1st}) also contains couplings proportional to $\hat F_z$
\begin{align}
\hat{H}^{||}_{\beta,\beta'} &= \Omega^{||}_{\beta,\beta'} \hat F_z \sin{(\Phi_{\beta,\beta'})}\\
\Omega^{||}_{\beta,\beta'} &= \frac{-g_F u |{\bf E}_{\beta} \times {\bf E}_{\beta'}|}{2\hbar g_S}\eta_{\beta,\beta'}\label{eq:omgparr}\\
\eta_{\beta,\beta'}& = \xibb\cdot {\bf e}_z,
\end{align}
where $\Omega^{||}_{\beta,\beta'}$ changes sign when $\xibb$ is aligned or anti-aligned with $B_{{\rm dc}}\ez$.  In the spin basis $\hat{H}^{||}_{\beta,\beta'}$ is simply a time dependent detuning; we shall explore a different set of basis states where $\hat{H}^{||}_{\beta,\beta'}$ produces an off-diagonal coupling.  

%The total effective coupling is simply
%%
%\begin{align}
%\hat H_{{\rm eff}} = \sum_{\beta\ne\beta'} \left[\hat{H}^{\perp}_{\beta,\beta'}+ \hat{H}^{||}_{\beta,%\beta'}\right]\label{eq:Heff2nd}
%\end{align}
%%

\subsection{Construction of fully coupled basis states}

For the remainder of this manuscript we narrow our discussion to the $f=1$ ground hyperfine manifold of $\Rb87$ and adopt the simplified labels $\ket{m_F}$, where $m_F\in\{-1,0,+1\}$ label hyperfine (spin) projections and $E_{m_F}$ label spin eigenenergies.  We divide the overall Zeeman shift into a scalar part which we neglect, a linear part given by $\hbar\delta_Z = (E_{-1}-E_{+1})/ 2$ and a quadratic part given by $\hbar\epsilon = (2E_0-E_{-1}-E_{+1})$.

We introduce the $\ket{X,Y,Z} = \ket{X},\ket{Y}$ and $\ket{Z}$ eigenstates, which consist of linear combinations of $\ket{m_F}$ states in the $f=1$ hyperfine manifold 
\begin{align}
\ket{X}=\frac{\ket{+1}-\ket{-1}}{\sqrt{2}}\text{, }\ket{Y}=i \frac{\ket{+1}+\ket{-1}}{\sqrt{2}}\text{, and }\ket{Z}=\ket{0}
\end{align}
The $\ket{X,Y,Z}$ state obey
\begin{align}
\frac{\epsilon_{jlm} \hat{F}_j}{\hbar} \ket{l}= i\ket{m}\label{eq:XYZ}
\end{align}
for indices $j,l,m$ in $\{x,y,z\}$.  The Raman couplings from the previous subsection have a spin dependence $\propto \hat F_x\text{, } \hat F_y\text{, or }\hat F_z$ and may therefore couple any pair of $\ket{X,Y,Z}$ states.  This observation was made recently by Ref.~\cite{Cooper2013} in the context of producing optical flux lattices.

A set of atomic eigenstates which approach the XYZ states can be produced by an oscillating magnetic field ${\bf B}_{\trf} \cos{(\omega_{\trf}+\gamma_{\trf})}$ that is orthogonal to $B_{dc}\ez$.  The rf coupling is described by 
\begin{align}
\hat H_{\trf} &= \frac{g_F \mu_B B_{\trf}}{\hbar}\cos{(\omega_{\trf}t+\gamma_{\trf})}(\xirf \cdot \hat{{\bf F}})
\end{align}
where
\begin{align}
\xirf &= \frac{{\bf B}_{\trf}}{|{\bf B}_{\trf}|}.
\end{align}
The rf is resonant with the average of the transitions that link adjacent hyperfine ground states, $\hbar \omega_{\trf} = \hbar\delta_Z$.  In the rotating frame of the rf and applying the RWA the $i\omega_{\trf} t$ term should be absent.  Likewise, the laser phases can generally be absorbed into the states.  The following Hamiltonian describes the rf coupling of the ground hyperfine states
\begin{align}
\hat{H}_{B} = (\delta_Z-\omega_{\trf})\hat{F}_z + \frac{\epsilon}{\hbar}(\hat{1}-\hat{F}_z^2)+\Omega_{\trf}\hat F_+e^{-i(\omega_{\trf}t+\gamma_{\trf})}+\text{h.c.}\label{eq:HB}
\end{align}
where $\Omega_{\trf} = g_F\mu_B B_{\trf}\xirf\cdot (\ex+i\ey)/2\hbar$.

\begin{figure}
\begin{centering}
\includegraphics[width=6in]{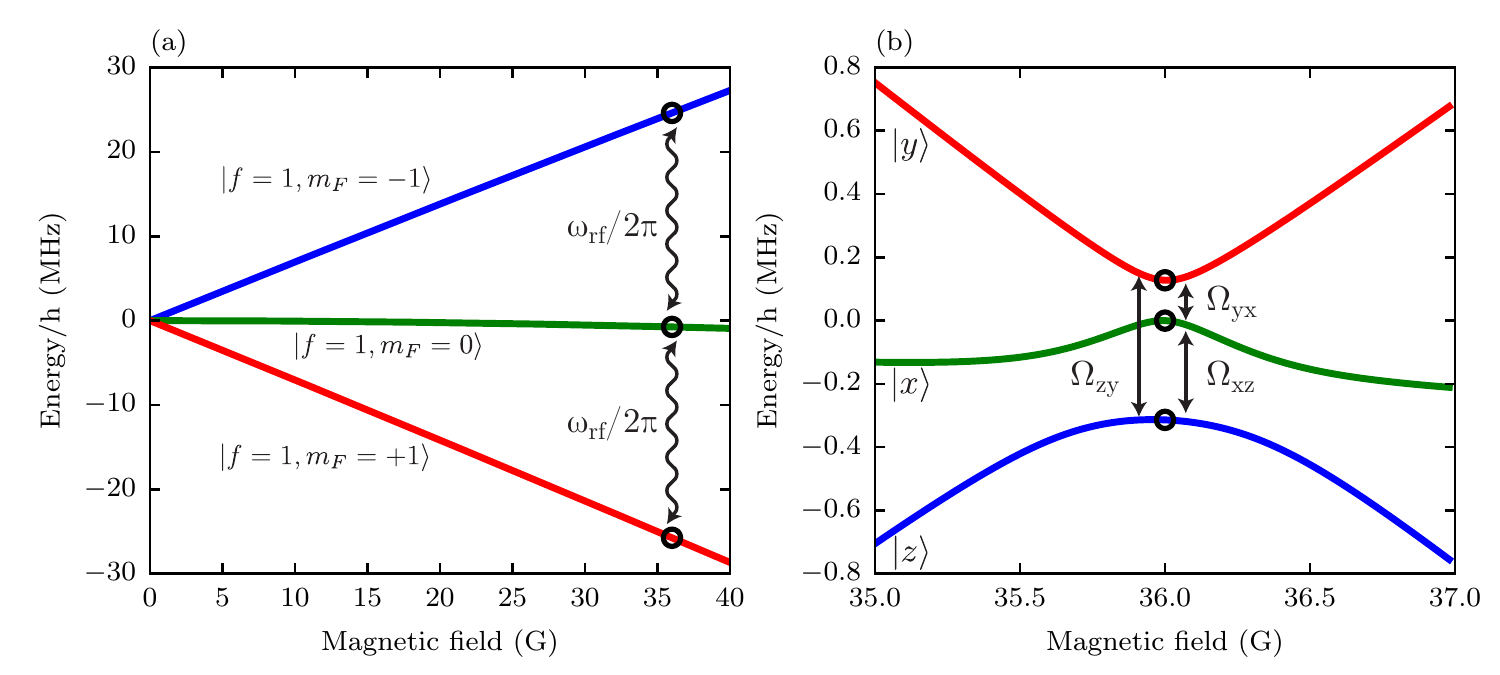}
\caption[frequencies and laser scheme]{\label{fig:XYZ} (a) Breit-Rabi calculation of the energy dependence of the magnetically sensitive spin-1 states.  A rf magnetic field with amplitude $1$ Gauss (G) applied within the $\ex-\ey$ plane links $\ket{m_F}$ states split by a dc field $B_{\text{dc}}=36\gauss$.  (b) In the frame rotating with the oscillating magnetic field the coupling opens gaps between the rf eigenstates, $\ket{x}\text{, }\ket{y}\text{, and }\ket{z}$.  These eigenstates are linked with resonant $\Omega_{y,x}\text{, }\Omega_{z,y}\text{, and }\Omega_{x,z}\text{+h.c.}$ Raman coupling.}
\end{centering}
\end{figure}

The rf eigenenergies $E_j$ of the Hamiltonian in the presence of the rf magnetic field are plotted verses ${\bf B}_{{\rm dc}}$ in Fig.~(\ref{fig:XYZ})b.  When we represent our rf eigenstates $\ket{x,y,z}$
\begin{align}
\ket{x}=&\ket{X}\xrightarrow{\epsilon/\Omega\rightarrow -\infty} \ket{X}\nonumber\\
\ket{y}=&\frac{-i2\arf\ket{Y}+\epsilon+\Omega_*\ket{Z}}{\sqrt{2}\Omega_*\sqrt{1+\frac{|\epsilon|}{\Omega_*}}}\xrightarrow{\epsilon/\Omega\rightarrow -\infty} \ket{Y}\nonumber\\
\ket{z}=&\frac{-i2\arf\ket{Y}+\epsilon-\Omega_*\ket{Z}}{\sqrt{2}\Omega_*\sqrt{1-\frac{|\epsilon|}{\Omega_*}}}\xrightarrow{\epsilon/\Omega\rightarrow -\infty} \ket{Z}\nonumber.
\end{align}
we see that they adiabatically approcah the $\ket{X,Y,Z}$ states as $\epsilon/\Omega\rightarrow -\infty$.
Here we defined $\Omega_{*}=\sqrt{\epsilon^2+4\arf^2}$.  

We resonantly link these eigenstates with the Raman coupling of the form described in the previous subsection and operate in the limit where the Raman coupling is much smaller than the rf coupling, $\Omega\ll\Omega_{\trf}$.  We define a rf eigenstate coupling matrix
\begin{align}
\hat D^l &=\sum_{j,j'} \ket{j}\bra{j}D^{l}_{j,j'}\ket{j'}\bra{j'}\label{eq:D}
\end{align}
where the matrix elements are
\begin{align}
D^{l}_{j,j'} &= \bra{j}\hat{F}_{l}\ket{j'}\\
\end{align}
which gives the representation of $\hat {\bf F}$ in the rf eigenbasis.   The $\hat D^x$ and $\hat D^y$ terms may be transformed into one another by changing the rf phase in Eq.~(\ref{eq:HB}): we choose the phases $\gamma_{\trf}, i\ln{(\xirf)}=0$ while defining the matrix elements, and we incorporate the rf phases into the definition of the total coupling in the next section.  These rf phases ultimately cancel in our coupling scheme.

The matrix elements $\bra{j}\hat{D}^{l}\ket{j'}$ of $\hat{D}^{l}$, linking rf-eigenstate pairs are
\begin{align}
\bra{x}\hat{D}^{y}\ket{z}=&i\hbar\sqrt{1-(\epsilon/\Omega_*)}\label{eq:Dy}\\
\bra{y}\hat{D}^{z}\ket{x}=&\frac{2\hbar(\arf/\Omega_*)}{\sqrt{1+(\epsilon/\Omega_*)}}\label{eq:Dz}\\
\bra{z}\hat{D}^{x}\ket{y}=&\sqrt{2}\hbar(\epsilon/\Omega_*)\label{eq:Dx}.
\end{align}
We can transform between the rf-eigenbasis and the $m_F$ basis using the rf-eigenstate coupling matrices, e.g. $\hat F_x\rightarrow \hat D^x$.

\subsection{Numerically calculating the eigenstates of the Raman and rf coupling}
\label{Floquet}

The rf and Raman couplings produce a time-periodic effective Hamiltonian.  Using Floquet theory we decompose the states of our Hamiltonian
\begin{align}
\ket{\psi(t)} = \sum_{n}c_{n}\ket{\psi(t)}_n = \sum_{n}c_n\exp{(-i\epsilon_n t/\hbar)}\ket{\phi(t)}_n
\end{align}
where $\epsilon_n = h n/T$ corresponds to the energy spacing between Floquet states when a time periodicity of $T$ exists in the Hamiltonian.

%For constant wave (CW) optical illumination the Floquet states are $\ket{\psi(t)}_{n} = \exp{(-i n \omega t )}\ket{\phi}_{n}$ where $n$ is the set of integers.  Because the CW Hamiltonian can be exactly decomposed into complex exponentials, $\ket{\phi}_n$ is time independent.  We apply a rotating frame transformation to each Floquet state, contributing an extra term $n\hbar\omega\ket{\psi}_n\bra{\psi}_n$ and making the Floquet Hamiltonian time independent.

\begin{figure}
\begin{centering}
\includegraphics[width=3in]{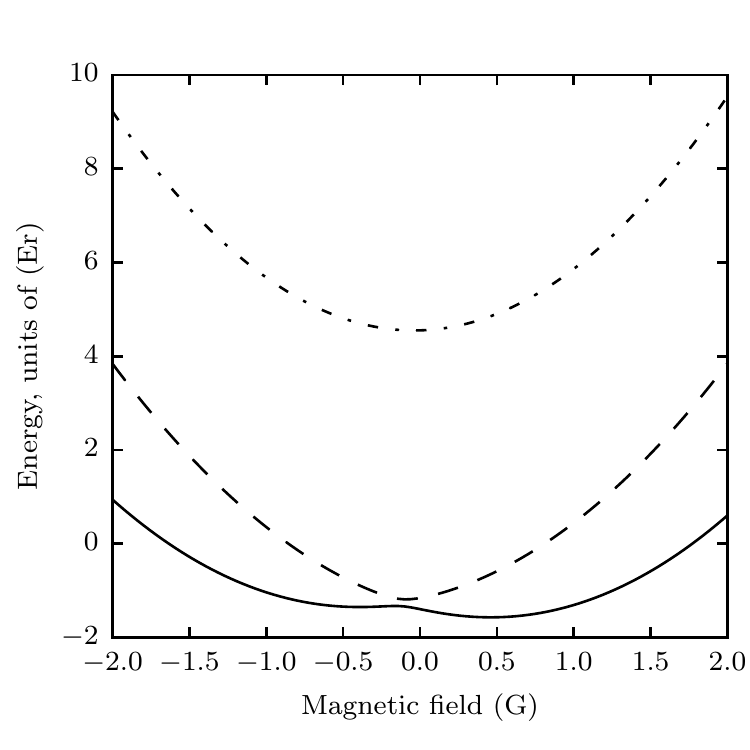}
\caption[Floquet Bands]{\label{fig:floquet} Calculated cross-sections along $q_x$ for Floquet bands.   $\hbar\arf=65\Er$, $\hbar\epsilon = -54\Er$, $\bar{\phi}=0$, $|\Omega_{x,z}|=|\Omega_{y,x}|=|\Omega_{y,z}| = \Omega$, $\hbar\Omega = 2\Er$, single-photon recoil is $\kr$ and all Raman beams are perpendicular.  These quantities are defined in the preceding sections.  The least (solid) and next least (dashed) energetic states are closely spaced while the most energetic (dashed dotted) state is separated by $3\hbar\Omega$ when $\bar{\phi}=0$.  The point at which the ground Floquet states cross is slightly displaced from the $q_x$ axis at these couplings due to the presence of nearby Floquet states.}
\end{centering}
\end{figure}

The Raman-rf CW Hamiltonian has multiple time periodicities and we use the RWA to eliminate rf and Raman coupling terms that very weakly couple the Floquet states.  We consider the parameter regime where $\Omega_{\trf}\gg \Omega$ and as a result we exactly diagonalize the ground hyperfine manifold with rf coupling and expand in terms of the Raman coupling.  The resulting Floquet Hamiltonian is
\begin{align}
\hat{H}_{{\rm Fl.}} =\sum_{n,m} \bigg\{&\left[\hat{H}_0+(n\hbar \omega_{x,z}+m\hbar\omega_{y,x})\hat{1}\right]\delta_{n,n'}\delta_{m,m'}\\
+&\bigg[\Omega_{x,z}^{\perp}\hat{D}^{\perp}\delta_{n-1,n'}\delta_{m,m'}e^{ -i(\gamma_{x,z}-\gamma_{\trf})}\nonumber\\
+&\frac{-i\Omega_{y,x}^{||}}{2}\hat{D}^{||}\delta_{n,n'}\delta_{m-1,m'}e^{-i\gamma_{y,x}}\nonumber\\
+&\Omega_{y,z}^{\perp}\hat{D}^{\perp}\delta_{n-1,n'}\delta_{m-1,m'}e^{ -i(\gamma_{y,z}-\gamma_{\trf})}\bigg]+\text{h.c.}\bigg\}\nonumber
\end{align}
where $\hat{1}$ is the identity in the rf-eigenbasis and the operators $\hat{H}_0$, $\hat{D}^{\perp} = (\hat{D}^{x}+i\hat{D}^{y})/2$ and $\hat{D}^{||}=\hat{D}^{z}$ are the $3\times 3$ matrices of rf eigenstates computed in the previous subsection. These are
\begin{align}
\hat{H}_0=\begin{pmatrix}
\left(\frac{\hbar^2(\boldsymbol{q}- \boldsymbol{K}_{y})^2}{2m}+E_{y}\right) & 0 & 0\\
0 & \left(\frac{\hbar^2(\boldsymbol{q}-\boldsymbol{K}_{x})^2}{2m}+E_{x}\right) & 0\\
0 & 0 & \left(\frac{\hbar^2(\boldsymbol{q}-\boldsymbol{K}_{z})^2}{2m}+E_{z}\right) \\
\end{pmatrix}
\end{align}
\begin{align}
\hat{D}^{\perp}=\hbar\begin{pmatrix}
4(\Omega_{\trf}/\Omega_*) & i\sqrt{1+(\epsilon/\Omega_*)} & \sqrt{2}(\epsilon/\Omega_*)\\
-i\sqrt{1+(\epsilon/\Omega_*)} & 0 & i\sqrt{1-(\epsilon/\Omega_*)}\\
\sqrt{2}(\epsilon/\Omega_*) & -i\sqrt{1-(\epsilon/\Omega_*)} & -4(\Omega_{\trf}/\Omega_*)\\
\end{pmatrix}
\end{align}
\begin{align}
\hat{D}^{||}=\hbar\begin{pmatrix}
0 & \frac{2(\Omega_{\trf}/\Omega_{*})}{\sqrt{1+(\epsilon/\Omega_{*})}} & 0\\
\frac{2(\Omega_{\trf}/\Omega_{*})}{\sqrt{1+(\epsilon/\Omega_{*})}} & 0 & \frac{1}{\sqrt{2}}\sqrt{1-(\epsilon/\Omega_{*})}\\
0 & \frac{1}{\sqrt{2}}\sqrt{1-(\epsilon/\Omega_{*})} & 0\\
\end{pmatrix}.
\end{align}
Figure~(\ref{fig:floquet}) depicts a quasi-energy unit cell.  When the Raman coupling coefficients exceed $2\Er$ (for our parameters) the location of the crossing point between the ground Floquet states drifts due to the presence of nearby Floquet states.  A changing location of this crossing point is also characteristic of imbalances in the Raman matrix elements that couple between rf eigenstates.  Adjusting the balance of laser intensities can return the crossing point to the origin where the degeneracy of the dispersion is maximized.

\subsection{Construction of a $\boldsymbol{3\times 3}$ Hamiltonian with fully coupled basis states}

In this section we apply the RWA to truncate the Floquet Hamiltonian at a single closed set of resonant couplings and obtain an effective $3\times 3$ Hamiltonian.  The validity of the RWA used to produce this Hamiltonian is determined by performing the numerics outlined in the previous subsection.  From this Hamiltonian, we analytically determine the conditions necessary to modify the gap between the ground Raman eigenstates.  

We take the form of the Raman coupling in Eqs.~(\ref{eq:omgperp}, \ref{eq:omgparr}) and transform them into the rotating frame of the rf, with angular frequency $\omega_{\trf}$.  Then we substitute the rf eigenstate coupling matrices from Eq.~(\ref{eq:D}) to determine the form of the Raman coupling in the rf eigenbasis:
\begin{align}
\hat{H}_{{\rm eff}} = \sum_{\beta\ne\beta'}&\bigg(\big[\Omega_{\beta,\beta'}^{\perp}\frac{\hat{D}^{x} + i\hat{D}^{y}}{2}\exp{(i[\Phi_{\beta,\beta'}+\omega_{\trf}t])}\theta(\omega_{\beta,\beta'})+\text{h.c.}\big]\\
&\Omega^{||}_{\beta,\beta'}\hat{D}^{z}\sin{(\Phi_{\beta,\beta'})}\bigg)\nonumber.
\end{align}
We require that the Raman frequency differences resonantly couple rf eigenstates $E_j$
\begin{align}
&\hbar\omega_{j,j'}\pm\hbar\omega_{\trf}(\delta_{j,z}-\delta_{j',z})=E_j-E_{j'}\label{eq:rfresonance}\\
&\hbar\omega_{j} = E_j\pm\hbar \omega_{\trf}\delta_{j,z}\label{eq:rfresonance2}.
\end{align}
The upper (lower) sign choice corresponds to blue (red) detuning.  This RWA is justified in the limit that $\Omega_{j,j'}\ll \omega_{j,j'}$ where $\omega_{j,j'} \approx \Omega_{\trf}$.

Using the laser polarizations recommended in the previous section, essentially setting $\boldsymbol{\xi}_{j,j'}$ parallel to ${\bf B}_{{\rm dc}}$ for coupling between $\ket{y}$ and $\ket{x}$ and perpendicular otherwise,  maximizes the ratio of couping to laser intensity.  The resonant terms comprise an effective Hamiltonian
\begin{align}
\hat{H}_{{\rm eff}} &= \sum_{j j'} \left\{\left[\frac{\hbar^2 {\bf k}^2}{2m}+E_j\right]\delta_{j,j'}+\left[\hbar \Omega_{j,j'}\exp{(i\Phi_{j,j'}^{\trf})}(1-\delta_{j,j'})+\text{h.c.}\right]\right\}\ket{j}\bra{j'}\label{eq:Heff_final}\\
\Omega_{z,y} &= \frac{\pm\Omega^{\perp}_{z,y}}{2} D^{x}_{z,y}\\
\Omega_{y,x} &= \frac{\mp i\Omega^{||}_{y,x}}{2} D^{z}_{y,x}\\
\Omega_{x,z} &= \frac{\pm i\Omega^{\perp}_{x,z}}{2} D^{y}_{x,z}
\end{align}
where $\Phi_{j,j'}^{\trf} = \Phi_{j,j'}\mp \hbar \omega_{\trf} (\delta_{j,z}-\delta_{j',z})$ and $\Omega_{j,j'} = \Omega_{j',j}^*$.  The upper (lower) sign choice corresponds to blue (red) detuning.  Following a unitary transformation, the Hamiltonian in Eq.~(\ref{eq:Hq}) is recovered.

From the matrix elements derived above, we may determine the phase sum $\bar \phi = \phi_{z,y}+\phi_{y,x}+\phi_{x,z}$ that contributes in Eq.~(\ref{eq:Energy}) to the overall energy of the combined Raman and rf eigenstates:
\begin{align}
\phi_{j,j'} =&i\ln{\left(\frac{\Omega_{j,j'}}{|\Omega_{j,j'}|}\right)}\\
\phi_{z,y} =& \mp\frac{\pi}{2}+\frac{\pi}{2}[1-\text{sign}{(u)}]+i\ln{\left(\frac{\boldsymbol{\xi}_{z,y}\cdot (\ex-i\ey)}{|\boldsymbol{\xi}_{z,y}\cdot (\ex-i\ey)|}\right)}\\
\phi_{y,x} =& \mp\frac{\pi}{2}+\frac{\pi}{2}[1-\text{sign}{(u)}]+\frac{\pi}{2}[1-\text{sign}{(\eta_{y,x})}]\\
\phi_{x,z} =& \pm\frac{\pi}{2}+\frac{\pi}{2}[1-\text{sign}{(u)}]+i\ln{\left(\frac{\boldsymbol{\xi}_{x,z}\cdot (\ex+i\ey)}{|\boldsymbol{\xi}_{x,z}\cdot (\ex+i\ey)|}\right)}\\
\bar \phi =& \mp \frac{\pi}{2}+\frac{\pi}{2}[1-\text{sign}{(u)}]+i\ln{\left(\frac{\boldsymbol{\xi}_{z,y}\cdot (\ex-i\ey)}{|\boldsymbol{\xi}_{z,y}\cdot (\ex-i\ey)|}\right)}\label{eq:barphi}\\
&+i\ln{\left(\frac{\boldsymbol{\xi}_{x,z}\cdot (\ex+i\ey)}{|\boldsymbol{\xi}_{x,z}\cdot (\ex+i\ey)|}\right)}+\frac{\pi}{2}[1-\text{sign}{(\eta_{y,x})}].\nonumber
\end{align}
We usually choose to make the two-photon matrix element $u$ negative: this contributes an overall factor of $\pi$ to $\bar \phi$.  Blue (red) detuning the Raman from the rf decreases (increases) $\bar{\phi}$ to $\pi/2$ ($3\pi/2$).  The two log terms in Eq.~(\ref{eq:barphi}) sum to a phase that is equivalent in radians to the azimuthal angle between the projections of $\boldsymbol{\xi}_{z,y}$ and $\boldsymbol{\xi}_{x,z}$ on the plane perpendicular ${\bf B}_{{\rm dc}}$.  The last term changes by a factor of $\pi$ when one log's argument changes sign.  Together, the last three terms on the RHS of Eq.~(\ref{eq:barphi}) contribute a phase to $\bar{\phi}$ bounded between $0$ and $\pi$.  The ground eigenstate of the effective Hamiltonian in Eq.~(\ref{eq:Heff_final}) is the Rashba potential when $\bar{\phi} = 0,2\pi$.  To produce this phase, the Raman must be red detuned from the rf (the lower sign choice) and the last three terms of Eq.~(\ref{eq:barphi}) must sum to $\pi/2$.  We describe a simple laser geometry in the appendix that satisfies this requirement.

\section{conclusion}

This proposal implements Rashba SOC using the ground atomic states of $\Rb87$.  As a result atoms cannot experience collisional deexcitation from the $f=2$ hyperfine manifold and the associated heating and decoherence that may disrupt many-body states.  Furthermore, we have exchanged technical challenges and expense associated with producing phase locked lasers separated by many GHz in frequency with the challenge of producing hundreds of $\kHz$ of rf coupling.  

% with part per thousand amplitude control  

\section{Acknowledgments}

This work was partially supported by the ARO's atomtronics MURI, by the AFOSR's Quantum Matter MURI, NIST, and the NSF through the PFC at the JQI.

\bibliography{RFRamanRashba}

\appendix*

\section{Appendix: Proposed preparation of experiment}

\subsection{Preparing the rf eigenstates}

We propose the application of $B_{{\rm dc}}$ along $\ez$ with an amplitude necessary to produce a $h\times 30\MHz$ linear Zeeman splitting between the ground hyperfine states of $\Rb87$.  In the presence of this magnetic field the quadratic Zeeman shift is $\approx h\times 250 \kHz$.  The ground hyperfine states are dressed by a $\omega_{\trf}/2\pi = 30 \MHz$ rf field with amplitude $\hbar\Omega_{\trf} = h\times 200\kHz$ that is set equal to the 2-photon resonance.  In $\Rb87$ the necessary amplitude of the rf magnetic field is $\sim 0.6\gauss$.  The polarization of ${\bf B}_{\trf}$ should be linear and orthogonal to $B_{{\rm dc}}$.

The Raman matrix elements given by Eqs.~(\ref{eq:Dy}, \ref{eq:Dx}, \ref{eq:Dz}) grow as $\epsilon/\arf\rightarrow -\infty$; the matrix element in Eq.~(\ref{eq:Dy}) is zero when $\epsilon=0$.  Simultaneously, as $\epsilon/\arf\rightarrow -\infty$ the gap between $\ket{x}$ and $\ket{y}$ closes
\begin{align}
\Delta^{{\rm rf}}_{y,x}=\frac{1}{2}\left(\epsilon+\sqrt{\epsilon^2+4\Omega_{\trf}^2}\right).
\end{align}
$\Delta^{{\rm rf}}_{y,x}$ is always the smallest gap in the system and $|\Delta^{{\rm rf}}_{y,x}|\rightarrow 0$ as $\epsilon/\arf\rightarrow -\infty$.  When $|\Delta^{{\rm rf}}_{y,x}|<|\Omega|$ the states $\ket{x}$ and $\ket{y}$ cannot be separately Raman coupled.  Similarly, $D_{y,z}^x$ is always the smallest matrix element in the system and $|D_{y,z}^x|\rightarrow 0$ as $\epsilon/\arf\rightarrow 0$.  We compute that the product $|D_{y,z}^x \Delta^{{\rm rf}}_{y,x}|$ is maximized when $-0.6<\epsilon/\arf < -1.1$.  

The ground eigenstate of the combined Raman and rf coupling becomes ring-like when the Raman coupling exceeds a characteristic energy scale $2\Er$ where $\Er = \hbar^2\kr^2/2m$ and $\hbar \kr$ is the single-photon recoil.  $2\Er$ is the kinetic energy gained when the Raman coupling mediates a spin flip and can vary between 0 and $4\hbar \kr/2m$ depending upon the laser geometry; $2\Er$ is based on a laser geometry where all the lasers are perpendicular to one another.  To produce the Rashba potential using our laser scheme and laser geometry the Raman coupling strength is bounded $2\Er<\hbar\Omega\ll \hbar\Delta^{{\rm rf}}_{y,x}$.

%We operate in the regime where the Raman coupling strength is much smaller than this splitting $\Omega_{y,x}\ll \Delta^{{\rm rf}}_{y,z}$.  There is also a photon recoil energy scale $2\hbar^2 \kr^2/2m\approx h\times 7.357\kHz\ll \Delta^{{\rm rf}}_{y,x}$ (the exact value is specific to our laser scheme) that must also be much smaller than the splitting between rf eigenvalues.

In alkali atoms, dc magnetic field fluctations often limit the long-term stability of an experiment that optically couples two or more magnetically split internal states.  This is partly the case because the splitting between internal states is nominally linear with magnetic field.  At resonance, the rf eigenstates respond quadratically to magnetic field fluctuations: $\Delta E_j \approx (g_F\mu_B \Delta B)^2/2\hbar \arf$.  When the rf coupling is strong $\hbar \arf=h\times 200\kHz$ compared to the Zeeman splitting amplitude fluctuations of a lab without active field control $~h\times 1\kHz$, the resulting impact of the magnetic fluctuations is reduced $\Delta E_j\approx h\times 5\Hz$.  Hence, rf eigenstates produced by sufficiently strong rf coupling become engineered clock states.  

%A coupling dependent shift in the location of the Dirac point (see the Floquet section) occurs when the Raman coupling is not $\hbar\Omega\ll \Delta_{y,x}$ or when $2\Er \ll \Delta_{y,x}$.  $2\Er$ is the typical energy spacing between the ground and first excited eigenenergies at the ground state minima.

\subsection{Raman laser frequencies, intensity and geometry}

\begin{figure}
\begin{centering}
\includegraphics[width=6in]{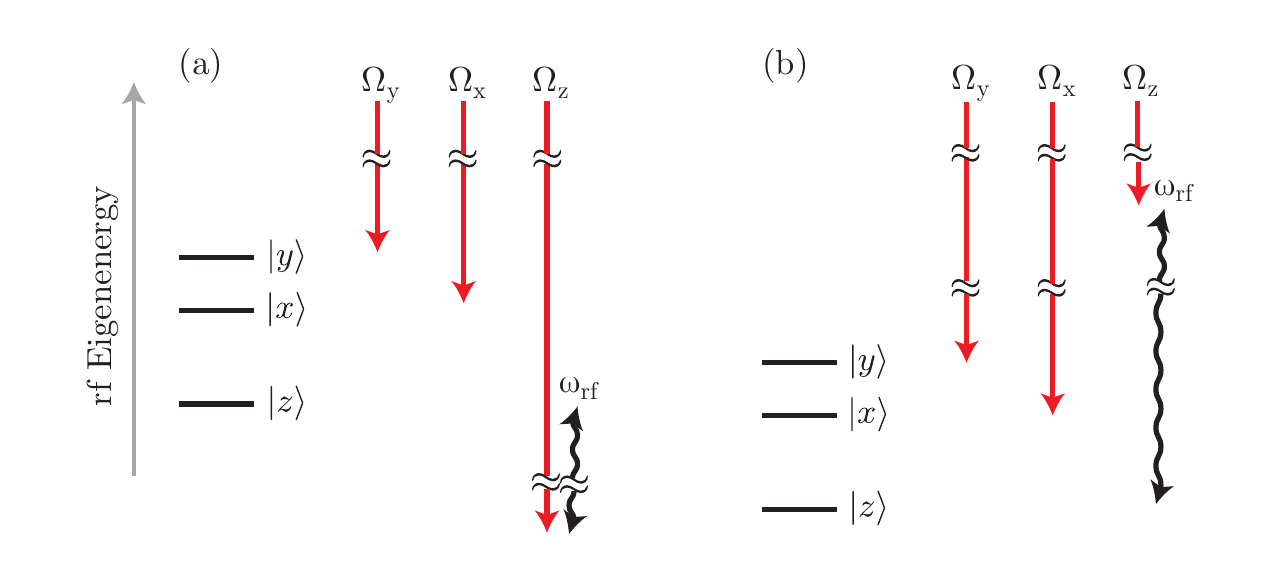}
\caption[frequencies and laser scheme]{\label{fig:freq} Level diagram for Raman coupling rf eigenstates.  (a) Here, $|\omega_{x,z}|,\,|\omega_{y,z}|>\omega_{\trf}$ and hence the Raman is blue detuned relative to the rf. (b) The Raman is red detuned relative to the rf.\label{fig:redblue}}
\end{centering}
\end{figure}

We illuminate a cloud of $\Rb87$ atoms using three linearly polarized lasers, all with wavelength very near $\lambda = 790.024\nm$.  At this wavelength, the 2-photon vector light shift matrix element $u$ is negative, while the scalar light shifts are zero.  The frequencies of these beams are
\begin{align}
(\omega_L-\omega_j)/2\pi = \omega_L \mp( \omega_{\trf}\delta_{j,z}\mp E_j/\hbar)\label{eq:freqs}
\end{align}
where $\omega_L = 2\pi c/\lambda$.  As shown in Fig.~(\ref{fig:freq}), the upper branch of Eq.~(\ref{eq:freqs}) corresponds to Raman frequency differences $\omega_{y,z},\omega_{x,z}>\omega_{\trf}$ while the lower branch switches the inequality.  Compared to the rf frequency the Raman coupling is blue and red detuned, respectively.

We write the Raman coupling in terms of the intensity of each laser 
\begin{align}
\Omega_{j,j'} &= \frac{\sqrt{I_{j} I_{j'}}}{R_{j,j'}I_0}\Omega\\
\Omega_0 &= \frac{g_F}{g_s}\frac{u I_0}{c \epsilon_0}\frac{1}{\hbar}
\end{align}
where $\Omega_0$ is an arbitrarily chosen coupling strength that we use as a benchmark and $R_{j,j'}$ is a dimensionless coefficient
\begin{align}
R_{z,y} &= \frac{4\hbar}{|\boldsymbol{\xi}_{z,y}\cdot (\ex-i\ey)||\hat D^{\perp}_{z,y}|}\\
R_{y,x} &= \frac{2\hbar}{|\boldsymbol{\xi}_{y,x}\cdot \ez||\hat D^{||}_{y,x}|}\\
R_{x,z} &= \frac{4\hbar}{|\boldsymbol{\xi}_{x,z}\cdot (\ex+i\ey)||\hat D^{\perp}_{x,z}|}
\end{align}
that compensates for laser geometry, applications of the RWA, and matrix elements.  We may then solve for the intensities in our system as a ratio of $I_0$ 
\begin{align}
\frac{I_{x}}{I_0} &= \frac{R_{x,y}R_{z,x}}{R_{y,z}}\\
\frac{I_{y}}{I_0} &= \frac{R_{y,z}R_{x,y}}{R_{z,x}}\\
\frac{I_{z}}{I_0} &= \frac{R_{z,x}R_{y,z}}{R_{x,y}}.
\end{align}
When $\epsilon/\arf = -0.8$ these ratios are $I_{x}/I_0=1.1$, $I_{y}/I_0=5.4$ and $I_{z}/I_0=21.5$.

\begin{figure}
\begin{centering}
\includegraphics[width=2in]{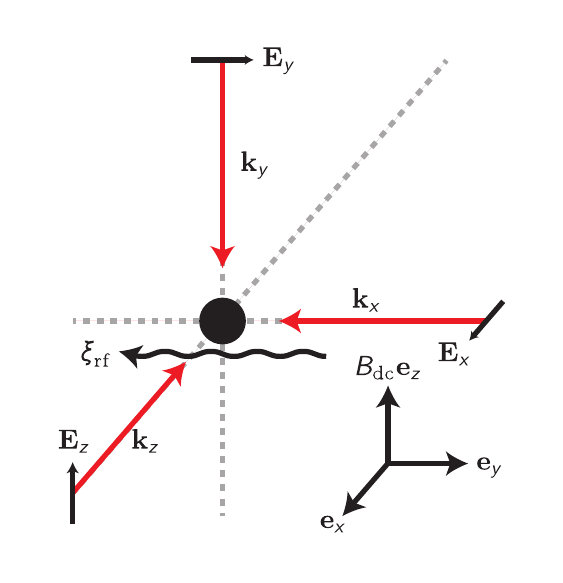}
\caption[frequencies and laser scheme]{\label{fig:geometry}  The wavevectors of all lasers are mutually perpendicular.  The polarizations (represented as the electric field at an instant in time) are also mutually perpendicular.  The $\pi$ polarized laser requires much more power and should be shifted in frequency by $\omega_{\trf}$ relative to the other two lasers.}
\end{centering}
\end{figure}

As shown in Fig.~(\ref{fig:geometry})a the wavevectors ${\bf k}_x$, ${\bf k}_y$, and ${\bf k}_z$ are aligned along $-\ey$, $-\ez$, and $-\ex$.  The electric fields ${\bf E}_x$, ${\bf E}_y$, and ${\bf E}_z$ of these lasers are polarized along $\ex$, $\ey$, and $\ez$.  The corresponding Raman coupling vector orientations are $\boldsymbol{\xi}_{z,y} = -\ex$, $\boldsymbol{\xi}_{x,z} = -\ey$, and $\boldsymbol{\xi}_{y,x} = -\ez$.  With red detuning and negative $u$, these parameters give $\bar \phi = 0$.

\end{document}